# Augmented reality as a Thirdspace: Simultaneous experience of the physical and virtual

Rob Eagle [0000-0001-8553-1713]

Digital Cultures Research Centre,
University of the West of England, Bristol, United Kingdom
`robert2.eagle@live.uwe.ac.uk`

**Abstract.** With the proliferation of devices that display augmented reality (AR), now is the time for scholars and practitioners to evaluate and engage critically with emerging applications of the medium. AR mediates the way users see their bodies, hear their environment and engage with places. Applied in various forms, including social media, e-commerce, gaming, enterprise and art, the medium facilitates a hybrid experience of physical and digital spaces. This article employs a model of real-and-imagined space from geographer Edward Soja to examine how the user of an AR app navigates the two intertwined spaces of physical and digital, experiencing what Soja calls a 'Thirdspace'. The article illustrates the potential for headset-based AR to engender such a Thirdspace through the author's practice-led research project, the installation *Through the Wardrobe*. This installation demonstrates how AR has the potential to shift the way that users view and interact with their world with artistic applications providing an opportunity to question assumptions of social norms, identity and uses of physical space.

## 1      Introduction

With nearly every smartphone capable of running applications that support augmented reality (AR) and with a number of AR 'smart' glasses presently in development, the spatial medium is fast approaching ubiquity (Azuma 2019). Applied in various forms, including social media, e-commerce, gaming and enterprise, AR can mediate the way users see their bodies, hear their environment and engage with places. The medium is already seeing widespread adoption and behaviour shifts through AR face filters on mobile platforms like Snapchat and Instagram and popular games like Pokémon GO. In shaping the user's experience of their environment through the illusion of real-time layered 3D imagery and audio, AR potentially presents the most radical shift in optics worldwide since Photoshop revolutionised digital image alteration three decades ago (Uricchio 2019).

   While AR was conceived as an ocularcentric medium that places digital objects or filters on top of (and responsive to) an environment, this paper argues for considering the medium as integrated *within* all the sensoriality of the physical world. According to geographer Edward Soja (1996), we can conceive of the physical and observable environment as a 'Firstspace', while the layer of imagination, images and media representations can be understood as a 'Secondspace'. The simultaneous experience and intertwining of those two spaces can create a *Third*space. In combining the physical environment with 3D imagery, spatial audio, haptics and more, AR is uniquely placed as a medium to facilitate a hybrid experience of the physical and virtual (Kosari and Amoori 2018).

To illustrate how engaging with the medium can engender a Thirdspace, I turn to findings from my own 3-year practice-based research project. *Through the Wardrobe* is an artistic interactive documentary in which the audience wears an AR headset to experience 3D animations and spatial audio combined with physical objects and scents in a room-scale installation. This creates an experience that does not exist only in the physical, non-digital space (Firstspace) or the headset on its own (Secondspace). In my use of the AR headset, movement and touch intertwine the Firstspace with the Secondspace, thereby creating a Thirdspace. *Through the Wardrobe* demonstrates how the medium can transform our understanding of and engagement with our bodies and space. The visitor is challenged to question their own labels, assumptions and gender expression through the intimate form of documentary storytelling via the headset and physical interaction with objects.

## 2      AR: a perceptual phenomenon, not a technology

While the term 'augmented reality' emerged from manufacturing applications in the aerospace industry in 1990 (Caudell & Mizell 1992), the medium was developed in a prototypical form in the 1960s (Billinghurst et al. 2015). Ivan Sutherland's 1968 head-mounted display (HMD), called the 'Sword of Damocles' – also recognised in the immersive industry as the progenitor of the modern VR headset – projected a 3D object onto half-silvered lenses in front of the eyes. When viewed through the headset, a computer-generated wireframe object appeared to hang in the air in the middle of the room. Engineering labs in the 1990s demonstrated multiple potential uses for these optical or 'see-through' HMDs, including as an aid for technicians (Feiner et al. 1993) and for training medical doctors (Bajura et al. 1992), but such headsets never saw widespread adoption in industry.

Artists in the 1970s and 80s (Krueger 1983) and engineers in the 90s developed alternative forms of AR that used a camera directed at a physical object, sending a video feed to a computer, which then generated real-time composited images overlaid with computer graphics. This is akin to the form of AR on smartphones and tablets with which most people today are familiar: a real-time image of a filter or 3D imagery composited on the image from the device's camera.

In an effort to define the medium based on the variety of applications and technologies used, Ronald T. Azuma (1997) identified three characteristics:
1. Combines real and virtual
2. Interactive in real-time
3. Registered in 3-D

Azuma's taxonomy establishes that it is not the device (e.g. PC monitor, handheld screen or HMD) or the technological output (optical or composite) that defines AR. It is the three characteristics via a digital interface that create the illusion of computer-generated elements appearing in the physical environment. As articulated by media theorist Horea Avram, AR can be understood as a 'perceptual phenomenon and not as a technology' (Avram 2016: 47); there is no single form or device that defines or restricts what the medium can be.

I would also emphasise here the importance of the role of the user in activating AR. The medium requires the bodily actions and movement of the user – whether that is holding a phone or wearing a headset or glasses – to make it interactive. Unlike a 'traditional' 2D medium, such as film, which can play linearly on a screen, the interactive quality of AR requires the physical presence and participation of the user to advance further. Additionally, while the story diegesis of a film takes place within the frame of the screen, an AR gives the illusion of digitally augmenting the user's body, home or a public space. In recent years, major fine artists have embraced apps like Acute and Snap to take work that would normally be seen only within a gallery to virtually any space in the world via a smartphone. The work of Olafur Eliasson, *WUNDERKAMMER* (2020) in the Acute Art app, for example, can bring AR elements like clouds or birds into the user's home. Such a use of AR opens the physical and financial accessibility of Eliasson's work to a wide audience, pushing beyond the limits of the rarefied art gallery.

For a game like Pokémon Go, players engage with the AR elements within the logic of the game world – the Pokémon characters that manifest in public spaces. This prolonged interaction with digital elements placed in physical spaces creates an intertwinement of realities for the players (Liberati 2018). The game creates a hybrid reality where physical actions (e.g. walking and tapping a phone) impact the digital game world (e.g. 'catching' a Pokémon), and the pursuit of the digital Pokémon (only visible via the phone screen) prompts physical actions (Liberati 2018: 218). Instead of producing a second world where the player is immersed (as in virtual reality), AR in a game like Pokémon Go shapes the player's perception of the physical environment, creating a mediated or hybrid experience of space.

With the proliferation of mobile devices since the 2000s, such as smartphones and tablets, artists, designers and engineers have evaluated ways of employing AR ubiquitously from indoor to outdoor settings (Schmalstieg and Reitmayr 2007). Now a number of wearable devices promise to take the medium from in-hand back to over-eye devices. Microsoft HoloLens (Fig. 1) and Magic Leap are presently the two leading see-through AR headsets in Europe and North America, but their technological limitations and high cost are factors in their lack of widespread consumer adoption (Azuma 2019). The world's largest tech companies, including Snap, Facebook, Apple and Google, are currently developing AR 'smart' glasses. There are also a handful of alternative AR headsets and glasses being released, such as Nreal in China. As these devices continue to develop, scholars and practitioners have the opportunity to evaluate and engage critically with emerging applications of AR.

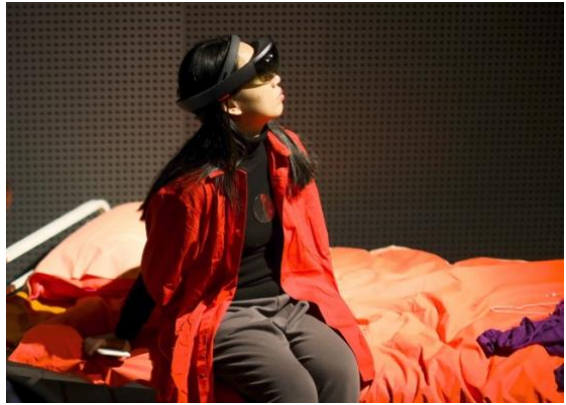

**Fig. 1.** The Microsoft HoloLens AR headset on a visitor to *Through the Wardrobe*. Credit: Goethe-Institut China (Beijing).

## 3  AR as a Thirdspace

According to geographer Edward Soja (1996), as critical studies underwent a 'spatial turn' in the 1980s and 90s, conceptualisations of space bifurcated into two approaches: 1) space conceived as concrete and material and 2) space as mental constructs and representations. In the first, we have the empirical, observable and mappable – what Soja terms 'Firstspace'. In the second, we have representations of space (images and imagination), theorised as simulation and simulacra. According to Soja, with the proliferation of representations and ideas through mass media, multiple and contested narratives now construct our understanding of the postmodern world. These do not take place in the abstract but occur within physical spaces. Soja proposes what he terms Thirdspace as a model for understanding spaces wherein the real and imagined simultaneously comingle:

> Everything comes together in Thirdspace: subjectivity and objectivity, the abstract and the concrete, the real and the imagined, the knowable and the unimaginable, the repetitive and the differential, structure and agency, mind and body, consciousness and the unconscious, the disciplined and the transdisciplinary, everyday life and unending history. (Soja 1996: 56-7)

Soja's Thirdspace provides a rich model for understanding the potential of AR in facilitating an experience with concurrent physical and representational realities. The user must orientate themselves in the Firstspace of the physical space (be it a room or an outdoor location) simultaneously within the Secondspace of the digital elements. Within this hybrid orientation of space, the user can pursue multiple interactions with different outcomes; there is not one single narrative. Players of Pokémon Go, for example, may pursue endless routes with a never-ending stream of Pokémon to catch.

For Soja, Thirdspace is more than a technological or mechanical mixing of the physical and representational. It is also a potentially *radical* space where multiple voices and representations of traditionally marginalised voices that exist beyond the dominant cultures of power can be expressed. Soja uses the example of a large postmodern urban complex in Los Angeles, called the Bonaventure, as a space in which multiple strata of society come together physically and in representations. Thirdspaces in which conflicting and disparate voices comingle allow the people within it to question social structures, such as race, class, gender and sexuality. Soja takes a Foucauldian stance on the politics of space: as all social spaces reflect and perpetuate social relations, we must see them as 'filled not only with authoritarian perils but with possibilities for community, resistance, and emancipatory change' (Soja 1996: 87). Perhaps an AR game like Pokémon Go may not at first seem like it has potential for radical change and resistance. Most AR apps, as officially developed by corporations, do not have an innate ability to 'emancipate' the marginalised and oppressed - unless those groups are enabled in hacking these technologies and developing them for themselves. AR *can* become a Thirdspace when the marginalised speak with and through the medium.

There are potential limits to applying Soja's model of Thirdspace to AR. Soja conceptualised Thirdspace in the mid-1990s as specific geographical sites. His examples focus mainly on urban sites, such as around Los Angeles, or borderlands between the US and Mexico. These are places where people live, and communities gather and contest the use of the land. For Soja, Thirdspace occurs in lived space and is sustained, like the Bonaventure complex. However, most AR experiences, such as games or social media face filters, are brief experiences. Applying the term Thirdspace to all uses of AR stretches Soja's original (potentially radical) intention for the model. However, that is not to say that we cannot adopt and modify the term to suit an analysis of AR, as others already have (Kosari and Amoori 2018). Perhaps AR is not a sustained Thirdspace at the moment, but as users increasingly engage with the medium in everyday lived environments, nearly anywhere with a smartphone or smart glasses can become a Thirdspace.

## 4 *Through the Wardrobe* as a Thirdspace

I present as a case study my own practice-as-research project, the AR installation *Through the Wardrobe*.[1] The installation was developed iteratively 2018–20, exhibited in documentary festivals and art spaces in the UK, Netherlands and China. When visitors enter the exhibition space, a host invites them to choose an item of clothing from the racks displayed by the entrance. All the clothing is secondhand and is either donated by or closely resembles the garments of four genderqueer and nonbinary people from Bristol, England: Micah, Jamie, Bec and Sammy. I audio interviewed these four contributors about their choice of clothing and its role in expressing gender identities that do not fit entirely into categories of male or female[2].

---

[1] This project has formed part of my PhD research within the Digital Cultures Research Centre, UWE Bristol. I am funded by the UK Arts & Humanities Research Council through the 3D3 Centre for Doctoral Training.

[2] For more on the distinction between genderqueer and nonbinary identities and expression, there are introductory information on the FAQ section of the project website: http://throughthewardrobe.net

Stories from the contributors examined everyday strategies for feeling strong, sexy or confident, regardless of acceptance within society. For example, Sammy was assigned male at birth, now uses she/her pronouns and is nonbinary. She speaks about how she uses accessories – a bow on her head, blusher on her cheeks and big silver hoop earrings – to give her a sense of feminine strength and energy that balances her beard stubble. Each contributor uses clothing, cosmetics or accessories to help physically express their identity as genderqueer.

In the installation, each item of clothing has a tag with a name and an icon in a circle (Fig. 2). Wearing their chosen item of clothing, the visitor then puts on the AR headset, the HoloLens, which scans the nametag and launches the story. This prompts an introduction from the owner of the clothing to begin in the headset, and the visitor then enters a room-scale installation resembling a bedroom. Approaching and interacting with each item of bedroom furniture triggers each chapter – five chapters in total – for each contributor. Instructions in the headset invite the visitor to sit or lay down, move through the room and try on additional items, such as boots or jewellery or even spray some perfume on the dressing table.

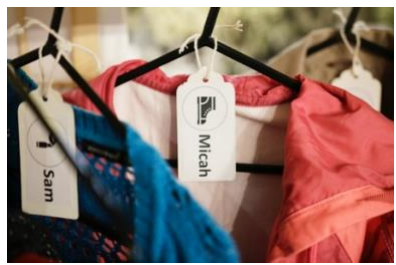

**Fig. 2.** Clothing tags with name and icon that is scanned by the headset to begin the story of the item's owner. Credit: Suzanne Zhang, Barbican Centre, London.

With most nonfiction AR experiences, digital elements are anchored to a flat surface, such as a tabletop, 2D picture or book page. In *Through the Wardrobe*, directional audio and 3D animations appear around the room and prompt the visitor to engage physically with the furniture and objects to reveal more of the story (Fig. 3). In Sammy's story, for example, text and animation in the headset invite the visitor to try something on from a selection of jewellery. When Bec talks about finding peace and quiet by sitting in an ancient stone circle, the visitor is invited to sit down in a chair, and 3D stone monoliths appear in a circle around the room (Fig. 4). This animation occurs only after the visitor has sat down. I wanted visitors to question the division of digital and physical; the physical experience depends on the augmented animations and instructions, and the virtual objects are triggered and engaged by physical objects.

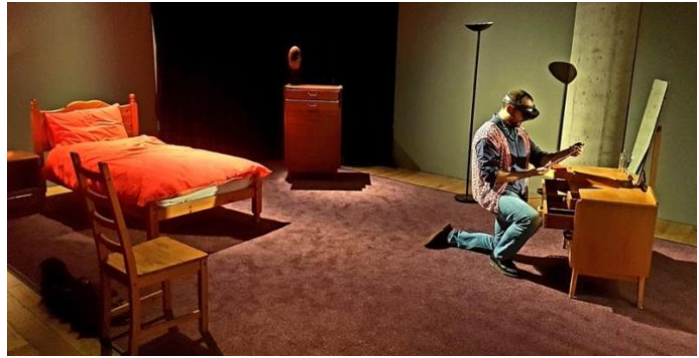

**Fig. 3.** Visitor to *Through the Wardrobe* engages with jewellery in Sammy's story, as exhibited at the gallery HOME, Manchester, UK.

Even though the visitor is wearing someone else's clothes and listening to their voice, the visitor is never playing a role or pretending to be anyone other than themself. Because the HoloLens is a see-through headset, the visitor can view their own physical body and environment with the addition of digital animations. The tone of *Through the Wardrobe* is intended to feel like an encounter with a friend in their bedroom. The four nonbinary contributors, who are often misunderstood or not recognised for their gender identity legally or socially, challenge the visitor to rethink Western cultural assumptions of the perceived binary division between male and female. The radical potential that nonbinary people present by not conforming to accepted legal or social frameworks align with Soja's model for Thirdspaces as spaces for radical challenge and change. As Thirdspaces centre marginalised voices, in *Through the Wardrobe*, only the voices of the four nonbinary contributors speak.

In the work, the AR device requires the choices and physical movement of the audience to progress the narrative. The audience are active agents in the unfolding of the experience. Their bodies navigate the Firstspace of the installation, trying on clothing and accessories, smelling perfume and sitting or lying down on the furniture. Simultaneously, they navigate the Secondspace of virtual content in the headset, physically walking through 3D animations, activating digital elements with their gaze and moving around spatial audio. The embodied choices of the visitor bring together the two spaces to create a Thirdspace of discovery and curiosity.

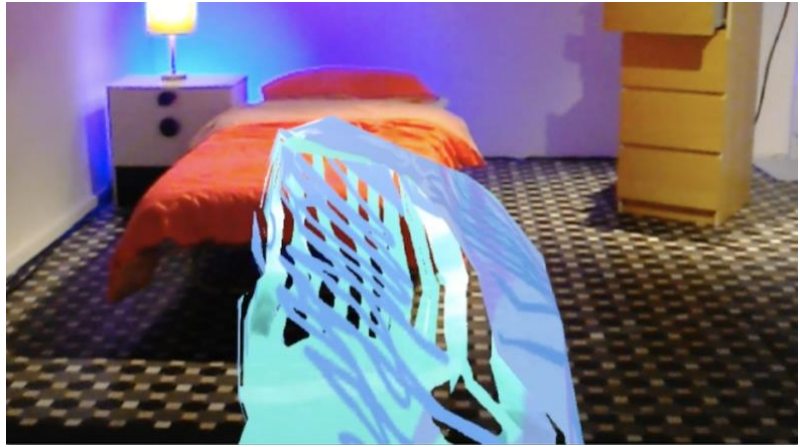

**Fig. 4.** In-headset view of animation of stone monolith rising from the floor, as exhibited at International Documentary Festival Amsterdam.

## 5 AR as a potentially radical medium for Thirdspace

With increasing applications of AR from composite forms in smartphones to optical forms projected in emerging headsets and smart glasses, the medium has the potential to shift the way users view and interact with the world. This change is already happening in face filters, e-commerce and gaming applications. But artists – and academic-practitioners – have the potential to use the medium for new forms of art and storytelling experiences. These can challenge audiences to question assumptions from gender identity to the limits of art galleries to gameplay in public spaces. AR invites the user to interact with both image and physical space through making creative choices.

*Through the Wardrobe* as an installation does not exist as a Thirdspace on its own, waiting to be entered, but rather is brought into being by the visitor's actions. What results is an experience unique for each visitor with variations on the four character stories, depending on with which physical and virtual materials they choose to engage. The immersant simultaneously experiences both the physical (such as a chair) with the virtual (where that chair becomes part of a stone circle). The physical and virtual intertwine as the visitor concurrently engages with both spaces. When marginalised voices speak with and through this blended reality, the visitor experiences the Thirdspace potential of AR.